\documentclass[%
 reprint,
showpacs,
 amsmath,amssymb,
 aps,
notitlepage,    
onecolumn,
showkeys
]{revtex4-1}

\usepackage{graphicx}
\usepackage{dcolumn}
\usepackage{bm}
\usepackage{subcaption}
\usepackage{hyperref}
\usepackage{slashed}
\usepackage[title]{appendix}
\usepackage{lipsum}

\usepackage{listings,color}
\definecolor{mygreen}{rgb}{0,0.6,0}
\definecolor{mygray}{rgb}{0.9,0.9,0.9}
\definecolor{mymauve}{rgb}{0.58,0,0.82}
\def\cnb#1{C$\nu$B}

\def\be{\begin{equation}}
\def\ee{\end{equation}}

\def\tu#1{\textup{#1}}

\def\bea{\begin{eqnarray}}
\def\eea{\end{eqnarray}}

\def\tr#1{\tu{tr}\left\{#1\right\}}

\def\math#1{{\itshape Mathematica }}
\def\fare#1{{\ttfamily FaRe}}
\def\fire#1{{\ttfamily FIRE}}

\def\algorithms{tarasov1,tarasov2,vermaseren,tarcer,air,fiesta,fire,mellinbarnes}
\def\reduction{tarasov3,tausk,anastasiou1,anastasiou2,anastasiou3}

\begin{document}

\preprint{??}

\title{\bf {\ttfamily FaRe}: a {\itshape Mathematica} package for tensor reduction of Feynman integrals}

\author{Michele Re Fiorentin}%
\email{m.re-fiorentin@soton.ac.uk}
\affiliation{%
 School of Physics and Astronomy\\
 University of Southampton\\
  SO17 1BJ, Southampton, U.K.
}%



\begin{abstract}
We present \fare{}, a package for {\itshape Mathematica} that implements the decomposition of a generic tensor Feynman integral, with arbitrary loop number, into scalar integrals in higher dimension. In order for \fare{} to work, the package {\ttfamily FeynCalc} is needed, so that the tensor structure of the different contributions is preserved and the obtained scalar integrals are grouped accordingly. \fare{} can prove particularly useful when it is preferable to handle Feynman integrals with free Lorentz indices and tensor reduction of high-order integrals is needed. This can then be achieved with several powerful existing tools. 
\end{abstract}

\maketitle


\section{Introduction}
The computation of Feynman integrals is a very active research field and many tools are able to perform reduction and computation (see for instance \cite{\algorithms}). Very often, these routines require numerators with loop momenta contracted with outer momenta. While this is not an obstacle to any integral computation, there may be cases in which explicit Lorentz indices are preferable or, even more in general, the tensor decomposition of a high-order tensor Feynman integral is required. Different ways to reduce a tensor integral to easier, known scalar ones are available, such as the well-known Passarino-Veltman reduction \cite{pave}, implemented, for example, in {\ttfamily LoopTools} \cite{looptools}. Nevertheless, the higher the order and the larger the number of outer legs, the harder the decomposition becomes. A reduction technique proposed in Refs. \cite{\reduction} appears very powerful and sufficiently machine-friendly to be implemented algorithmically. The package we propose, \fare{}, deals with Feynman integrals, with arbitrary number of loops. The propagators can appear to any power and the tensor integral to decompose can be any rank. The output generated will be a combination of scalar integrals, in higher dimension, with different powers of the propagators. Most importantly, the tensor structure is preserved, so that it is possible to identify all the different Lorentz contributions in the final result and use it for further calculations in {\itshape Mathematica}. 

\fare{} requires the package {\ttfamily FeynCalc} \cite{feyncalc} in order to perform the decomposition and supply the result. By means of a built-in function already implemented, it is possible to get an output suitable for further reduction to master integrals with \fire{} \cite{fire}. It is the usage of \fare{} in combination with other tools, such as \fire{}, that can prove to be rather useful in many different ways, even though the results provided by \fare{} alone are always self-consistent.

This paper is structured as follows: in Section~\ref{sec:algorithm} we give a brief overview of the algorithm, in Section~\ref{sec:package} we describe the package \fare{} and in Section~\ref{sec:conclusions} we draw the conclusions. Finally, in the Appendix an explanatory example is considered.
\section{The algorithm}
\label{sec:algorithm}
\fare{} implements the tensor reduction algorithm proposed by \cite{\reduction}, that makes use of raising and lowering operators. The procedure is briefly reviewed in what follows. 

As a starting point, we consider the $N$-loop scalar Feynman integral in $D$-dimensional Minkowski space
\begin{equation}
\label{eq:int1}
I^D_{\nu_1,\dots,\nu_n}\equiv\int\!\prod_{l=1}^N\frac{d^Dk_l}{i\pi^{D/2}}\frac{1}{A_1^{\nu_1}\cdots A_n^{\nu_n}}, 
\end{equation}
where $k_l$ are the loop momenta and $1/A_i$ is a propagator whose generic exponent is $\nu_i$. This integral can be re-written using the Schwinger parameterisation as
\begin{align}
\label{eq:int2}
I^D_{\nu_1,\dots,\nu_n}=&\int\!\prod_{l=1}^N\frac{d^Dk_l}{i\pi^{D/2}}\left(\int_0^\infty\!\prod_{i=1}^n\frac{(-1)^{\nu_i}}{\Gamma(\nu_i)}
\,x_i^{\nu_i-1}dx_i\right)\exp\left(\sum_{i=1}^nx_i\,A_i\right),
\end{align}
where, in general, we have
\begin{equation}
\label{eq:exp1}
\sum_{i=1}^nx_iA_i=\sum_{l=1}^Na_l\,k_l^2+2\sum_{l=1}^N\sum_{h=l}^N\,c_{lh}\left(k_h\cdot k_l\right)+2\sum_{l=1}^Nd_l\cdot k_l+f.
\end{equation}
Here $a_l$ and $c_{lh}$ are polynomial in $x_i$, while $d_l^\mu$ and $f$ are functions of both $x_i$ and the outer momenta. 
This expression can then be diagonalised by performing a momenta transformation. It is convenient to transform the loop momenta in sequence. Calling $\mathcal{C}^{\mu}(k_l)$ and $\mathcal{C}(k_l^2)$ the coefficients of the $l$-th momentum and of its square respectively, we can replace $k_l^\mu$ with the new momentum
\begin{equation}
\label{eq:trans1}
K^\mu_l=k_l^\mu-\frac{\mathcal{C}^\mu(k_l)}{2\,\mathcal{C}(k_l^2)}.
\end{equation}
It is clear that the first momentum to be transformed will have the trivial values $\mathcal{C}^\mu(k_1)=2d^\mu_1+2\sum_{j=1}^N c_{1j}\,k_j^\mu$ and $\mathcal{C}(k^2_1)=a_1$, but the following replacements will have a more complicated structure. Once all the $N$ replacements have been performed, the exponent will be
\begin{equation}
\label{eq:exp2}
\sum_{i=1}^nx_iA_i=\sum_{l=1}^N\mathcal{C}(k_l^2)K_l^2+\frac{Q}{P},
\end{equation}
where $Q/P$ regroups all the terms zero-order in all the loop momenta. The integration over the new loop momenta $K_l$'s can then be performed as standard gaussian integration, so that we end with
\begin{align}
\label{eq:int3}
I^D_{\nu_1,\dots,\nu_n}&=\left(\int_0^\infty\!\prod_{i=1}^n\frac{(-1)^{\nu_i}}{\Gamma(\nu_i)}
\,x_i^{\nu_i-1}dx_i\right)\frac{1}{P^{D/2}}\exp\left(\frac{Q}{P}\right)\nonumber\\
&\equiv\int\!\mathcal{D}x\,\mathcal{I}^D_{\nu_1,\dots,\nu_n},
\end{align}
where we defined
\begin{equation}
\int\!\mathcal{D}x\equiv\int_0^\infty\!\prod_{i=1}^n\frac{(-1)^{\nu_i}}{\Gamma(\nu_i)}.
\,x_i^{\nu_i-1}dx_i
\end{equation}

We can identify a tensor integral according to the momenta that appear in the numerator, by giving the list $\mathcal{N}$ of their labels. For instance, a tensor integral like
\begin{equation}
I^D_{\nu_1,\dots,\nu_n}(k_1^\mu k_1^\nu k_5^\rho)\equiv\int\!\prod_{l=1}^N\frac{d^Dk_l}{i\pi^{D/2}}\frac{k_1^\mu\,k_1^\nu\,k_5^\rho}{A_1^{\nu_1}\cdots A_n^{\nu_n}}, 
\end{equation}
can be specified by $\mathcal{N}=(1,1,5)$, so that we can refer to the elements of the list as $\mathcal{N}_j$ (e.g. $\mathcal{N}_1=1$, etc.). Clearly, the order in $\mathcal{N}$ is arbitrary. From eq.~(\ref{eq:int2}) and (\ref{eq:exp1}), a generic tensor integral of rank $\mathcal{R}$ can be obtained as
\begin{align}
\label{eq:tens1}
I^D_{\nu_1,\dots,\nu_n}&(\mathcal{N}^{\mu_1\cdots\mu_\mathcal{R}})=\int\!\prod_{l=1}^N\frac{d^Dk_l}{i\pi^{D/2}}\int\!\mathcal{D}x\,\prod_{j=1}^\mathcal{R}\frac{1}{2}\partial_{\mathcal{N}_j}^{\mu_j}\,\exp\left(\sum_{i=1}^nx_i\,A_i\right),
\end{align}
where we defined 
\begin{equation}
\partial_{\mathcal{N}_j}^{\mu_j}\equiv\frac{\partial}{\partial (d_{\mathcal{N}_j})_{\mu_j}}.
\end{equation}
Performing on eq.~(\ref{eq:tens1}) the diagonalisation and gaussian integration discussed above, we get
\begin{equation}
\label{eq:tens2}
I^D_{\nu_1,\dots,\nu_n}(\mathcal{N}^{\mu_1\cdots\mu_\mathcal{R}})=\int\!\mathcal{D}x\frac{1}{P^{D/2}}\prod_{j=1}^\mathcal{R}\frac{1}{2}\partial_{\mathcal{N}_j}^{\mu_j}\,\exp\left(\frac{Q}{P}\right).
\end{equation}
The action of the derivative operations on the exponential will result in a function $W^{\mu_1\cdots\mu_\mathcal{R}}(x_i;p_\tu{e})$ which contains all the allowed tensor structures that can be formed with the external momenta $p_e$, and whose coefficients are functions of the Schwinger parameters $x_i$. We have
\begin{equation}
\label{eq:tens3}
I^D_{\nu_1,\dots,\nu_n}(\mathcal{N}^{\mu_1\cdots\mu_\mathcal{R}})=\int\!\mathcal{D}x\,W^{\mu_1\cdots\mu_\mathcal{R}}(x_i;p_\tu{e})\,\mathcal{I}^D_{\nu_1,\dots,\nu_n}.
\end{equation}
All the coefficients in $W^{\mu_1\cdots\mu_\mathcal{R}}(x_i;p_\tu{e})$, polynomials in $x_i$, can be absorbed in the integration measure. Specifically, each $x_i$  gives a factor $-\nu_i$ and raises the relative exponent $\nu_i\rightarrow\nu_i+1$, while each factor $1/P$ modifies the dimension of the integral $D\rightarrow D+2$. Therefore, the tensor integral we started with is decomposed in a combination of scalar integrals in higher dimension, with different propagator exponents and regrouped according to the tensor structures given by the outer momenta. 

For detailed examples, we refer the reader to the literature (e.g. \cite{anastasiou3}).

\fare{} automatically implements this procedure with arbitrary loop number $N$ and tensor rank $\mathcal{R}$.
\section{The package}
\label{sec:package}
The package \fare{} can be downloaded at \url{https://sourceforge.net/projects/feyntoolfare/}.
It does not require any installation and it can be called in your {\itshape Mathematica} notebook simply by typing {\ttfamily <}{\ttfamily <FaRe.m}. It requires the package {\ttfamily FeynCalc}, which must be loaded in the notebook, lest \fare{} loading automatically aborts.

The functions available in \fare{} and their arguments are listed below.
\begin{lstlisting}[language=Mathematica,numbers=left,numbersep=8pt,numberstyle=\small\color{black}]
QP[ loops ]
TRed[ D,num,den,loopMomenta ]
FIREType[ expr,zeros,irr,fLetter ]
\end{lstlisting}

The function {\ttfamily QP} returns the form of $Q/P$ in eq.~(\ref{eq:tens2}), after the momentum transformations in eq.~(\ref{eq:trans1}). The quantity returned does not contain the zero-order term $f$ (see eq.~(\ref{eq:exp1})), which can be easily added by hand.  The output is in terms of coefficients $\mbox{\ttfamily d}_i^\mu$ and $\mbox{\ttfamily ac}_{ij}$, where $\mbox{\ttfamily ac}_{ii}=a_i$ and $\mbox{\ttfamily ac}_{ij}=c_{ij}$ for $i\neq j$. The only argument of the {\ttfamily QP} function is the number of loops. Here we give as an example the form of $Q/P$ in the two-loop case
\begin{lstlisting}[language=Mathematica,numbers=left,numbersep=8pt,numberstyle=\small\color{black}]
QP[ 2 ]
     (ac11 d2@$^2$@ -2 ac12 d1 d2 +ac22 d1@$^2$@)/(ac11 ac22 -ac12@$^2$@)
\end{lstlisting}
where we can see that $P=a_1a_2-c_{12}^2$ and $Q=a_1d_2^2-2c_{12}\,d_1\cdot d_2+a_2\,d_1^2$. Adding the $f$ term, we get $Q=a_1d_2^2-2c_{12}\,d_1\cdot d_2+a_2 d_1^2+f(c_{12}^2-a_1a_2)$, that matches the results found in the literature.

The function {\ttfamily TRed} is the core of the package \fare{}. It performs the tensor reduction of the integral provided. {\ttfamily TRed} has four arguments
\smallskip
\begin{lstlisting}
TRed[D,num,den,loopMomenta] 
\end{lstlisting}

\begin{itemize}
\item {\ttfamily D} is an integer representing the dimension $D$ in which the integral must be computed. Notice that, using dimensional regularisation, the regulator must be left implicit and not specified in $D$.
\item {\ttfamily num} is a list containing the loop momenta that appear in the numerator of the integral.
\item {\ttfamily den} is a list containing the arguments $\bar{A}_i$ of the propagators $A_i$, and their relative exponent $\nu_i$, in the form {\ttfamily \{\{$\bar{A}_1,\nu_1$\},\dots,\{$\bar{A}_n,\nu_n$\}\}}. Each $\bar{A}_i$ must be written without the square.
\item {\ttfamily loopMomenta} is a list containing the loop momenta.
\end{itemize}

The output is a combination of scalar integrals gathered by the tensor structures formed with the external momenta. Integration over the Schwinger parameters is understood. The tensor structures that appear in the output are {\ttfamily FeynCalc} objects that can be manipulated in further computations. Indices and vector names are {\itshape Mathematica} strings.

\begin{figure}[h]
\centering
\includegraphics[height=5cm]{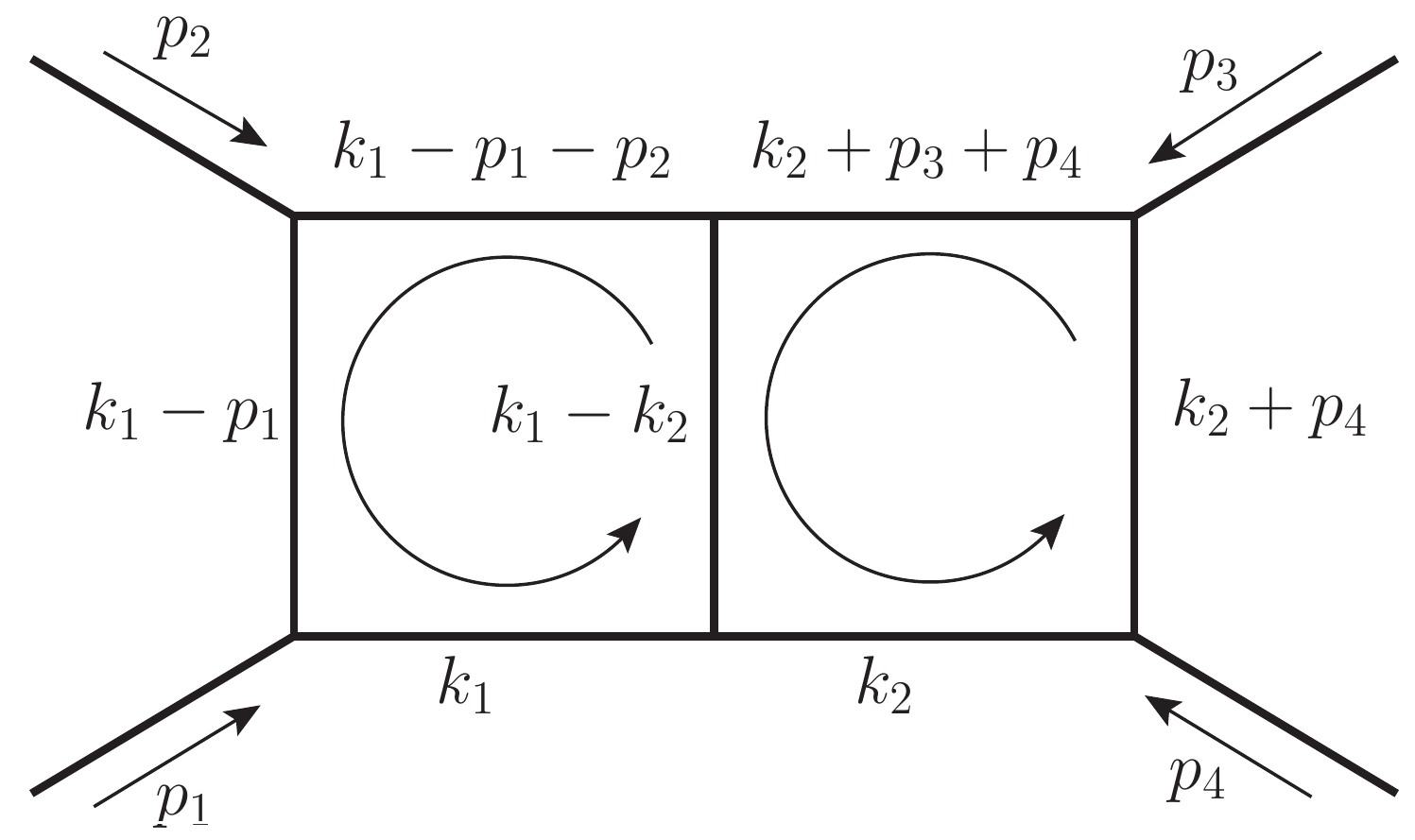}
\caption{External and loop momenta for the double-box integral.}
\label{fig:db}
\end{figure}

To illustrate how this function works, we can compute a double-box vector integral in three dimensions. Taking the momenta as in fig.~\ref{fig:db},
we can compute the vector integral
\begin{widetext}
\begin{align}
\label{eq:db1}
I^D_{(1,1,1,1,1,1,1)}(k_1^2\,k_1^\mu)&\equiv\int\!\frac{d^Dk_1}{i\pi^{D/2}}\frac{d^Dk_2}{i\pi^{D/2}}\frac{k_1^2\,k_1^\mu}{k_1^2\left(k_1-k_2\right)^2\left(k_1-p_{12}\right)^2\left(k_1-p_1\right)^2k_2^2\left(k_2+p_4\right)^2\left(k_2+p_{34}\right)^2}\\
=&\int\!\frac{d^Dk_1}{i\pi^{D/2}}\frac{d^Dk_2}{i\pi^{D/2}}\frac{k_1^\mu}{\left(k_1-k_2\right)^2\left(k_1-p_{12}\right)^2\left(k_1-p_1\right)^2k_2^2\left(k_2+p_4\right)^2\left(k_2+p_{34}\right)^2},
\label{eq:db2}
\end{align}
\end{widetext}
where $p_{12}=p_1+p_2$ and $p_{34}=p_3+p_4$. In the second step we have cancelled a propagator against $k_1^2$ in the numerator. The corresponding \fare{} input is
\begin{lstlisting}[numbers=left,numbersep=8pt,numberstyle=\small\color{black},firstnumber=1]
db1=TRed[3, {k1}, {{k1-k2,1},{k1-p12,1},{k1-p1,1},
	{k2,1},{k2+p4,1},{k2+p34,1}},  {k1,k2}] 
\end{lstlisting}
Here, only the vector arguments of the propagators appear in the list, together with the relative exponent, in this case all equal to 1. The outer momenta are automatically converted into {\ttfamily FeynCalc} four-vectors by \fare{} and treated accordingly. The output produced by \fare{} is
\smallskip
\begin{lstlisting}[numbers=left,numbersep=8pt,numberstyle=\small\color{black},firstnumber=3,basicstyle=\linespread{1.7}\ttfamily]
@-$p1^{\mu1}\left(\mathcal{I}^5_{\{1,1,1,1,2,2\}}+\mathcal{I}^5_{\{1,1,1,2,1,2\}}+\mathcal{I}^5_{\{1,1,2,1,1,2\}}\right)-p12^{\mu1}\left(\mathcal{I}^5_{\{1,2,1,1,2,1\}}+\mathcal{I}^5_{\{1,2,1,2,1,1\}}+\mathcal{I}^5_{\{1,2,2,1,1,1\}}\right)+$@
	@$p34^{\mu1}\,\mathcal{I}^5_{\{1,1,2,1,2,1\}}+p4^{\mu1}\,\mathcal{I}^5_{\{1,1,1,2,2,1\}}$@
\end{lstlisting}
It is convenient to point out that the four-vectors in the output are in the form
\smallskip
\begin{lstlisting}
@$p1^{\mu1}$@//InputForm = Pair[LorentzIndex["@$\mu$@1"], Momentum["p1"]]
\end{lstlisting}
where both the momentum name and its index are strings.

The output produced by {\ttfamily TRed} can be further manipulated in {\itshape Mathematica}, and it can be used as input to \fire{}, in order to reduce the obtained scalar integrals to master ones. To this aim, the function {\ttfamily FIREType} can be employed. This function has 4 arguments
\smallskip
\begin{lstlisting}
FIREType[expr,irr,zeros,fLetter]
\end{lstlisting}
\begin{itemize}
\item {\ttfamily expr} is a linear combination of scalar integrals as obtained from {\ttfamily TRed} output. It is convenient to remove the tensor structures, even if it is not strictly necessary.
\item {\ttfamily irr} is the number of irreducible numerators treated by \fire{}. Their exponents will be appended to the $\nu_i$'s lists and will appear as zeros.
\item {\ttfamily zeros} is the list of positions where to insert a zero in the $\nu_i$'s list. This can be used if \fire{} asks for a list of exponents longer than that used by \fare{}, for instance because of simplifications.
\item {\ttfamily fLetter} specifies the name used to represent the integrals in a suitable way for \fire{}.
\end{itemize}
Referring to the previous example, we can apply {\ttfamily FIREType} to the coefficient of $p_1^{\mu_1}$. \fire{} will accept the seven denominators as in eq.~(\ref{eq:db1}) plus two irreducible numerators, so that it will handle lists with nine elements. Due to the simplification in eq.~(\ref{eq:db2}) the list produced by {\ttfamily TRed} have only six indices while, at the same time, {\ttfamily TRed} assures we do not have numerators anymore. The input and output of {\ttfamily FIREType} are then
\smallskip
\begin{lstlisting}[numbers=left,numbersep=8pt,numberstyle=\small\color{black},firstnumber=5,basicstyle=\linespread{1.7}\ttfamily]
cp1=Coefficient[db1, Pair[LorentzIndex["@$\mu$@1"], Momentum["p1"]]
    @$-\left(\mathcal{I}^5_{\{1,1,1,1,2,2\}}+\mathcal{I}^5_{\{1,1,1,2,1,2\}}+\mathcal{I}^5_{\{1,1,2,1,1,2\}}\right)$@

FIREType[cp1,2,{3},G]
    -G[{1,1,0,1,1,2,2,0,0}] - G[{1,1,0,1,2,1,2,0,0}]
    -G[{1,1,0,2,1,1,2,0,0}]
\end{lstlisting}
where two more indices are added and a zero is inserted at position 3, due to the cancellation in eq.~(\ref{eq:db2}). In the last arguments we have specified that the integrals will have function name {\ttfamily G}. The output can be used as input to \fire{} once converted to expression. It must be noticed that {\ttfamily FIREType} does not specify the integral dimension, that must be separately taken care of.

An explanatory example on the usage of the functions available in \fare{} and, more specifically, on its employment with {\ttfamily FeynCalc} and \fire{}, can be found in the Appendix.
\section{Conclusions}
\label{sec:conclusions}
We have presented the new package for {\itshape Mathematica}, \fare{}, that automatises dimensional tensor reduction of an arbitrary-rank tensor Feynman integral with arbitrary number of loops. Its usage can prove to be useful for computation of high-order tensor Feynman diagrams, providing a rather easy way to obtain their tensor decomposition. The subsequent reduction of the resulting scalar integrals to master ones can be achieved by means of other powerful existing tools such as \fire{}. The possibility of interfacing \fare{} with \fire{} is made easier through the function {\ttfamily FIREType}. However, improvements can be made in this respect. 

The synergetic employment of \fare{}, \fire{} and, of course, {\ttfamily FeynCalc} can thus prove to be quite powerful in handling high-order Feynman integrals. In the Appendix we sketch a possible example, borrowed from ABJM theory \cite{abjm}, that can be find in its detailed version in the example notebook attached to the \fare{} distribution. Here we compute the a double-box integral in three dimensions. Though already well-known, the result obtained with \fare{} does not represent just a benchmark of the code, but shows hidden identities which may prove to be interesting (see for instance eq.~(\ref{eq:canc})).

The package \fare{} is fully open source, which represents one of its strong point. Open access to the source code allows for improvements from the whole scientific community. This way, the door stays always open for countless customisations that may help to tackle many different problems in high energy Physics.
\section*{Acknowledgements}
The author is particularly indebted to Setareh Fatemi and Amedeo Primo for the inspiration and their encouragement since the very early stages of this work. The author also wishes to thank Lorenzo Bianchi for several, very useful discussions and comments. 
MRF acknowledges financial support from the STAG Institute and is also deeply grateful to the Physics Department of the University of Torino for the kind hospitality.
\appendix
\section{A pedagogical example: double-box in three dimensions}
As an explanatory example of the functioning of \fare{} and its interplay with other reduction tools such as \fire{}, in this appendix we compute a double-box integral entering the planar two-loop ABJM amplitude \cite{chen,leoni,lore}. The full computation is reported in the example {\itshape Mathematica} notebook that can be found together with the package \fare{}.

 Defining the momenta as in fig.~\ref{fig:db}, we compute the integral
\begin{align}
\label{eq:dbp}
&\mathbf{DB}_P(s,t)\equiv\int\!\frac{d^Dk_1}{i\pi^{D/2}}\frac{d^Dk_2}{i\pi^{D/2}}\frac{\mathcal{N}_P}{t\,k_1^2\left(k_1-p_{12}\right)^2k_2^2\left(k_2+p_{34}\right)^2\left(k_2+p_4\right)^2\left(k_1-k_2\right)^2\left(k_1-p_1\right)^2},
\end{align}
where
\begin{align}
\label{eq:np}
\mathcal{N}_P=&\left[s\,\tr{k_1\,p_1p_4}+k_1^2\,\tr{p_1p_2p_4}\right]\left[s\,\tr{k_2\,p_1p_4}+k_2^2\,\tr{p_1p_2p_4}\right].
\end{align}
The integrals to be decomposed by \fare{} are just three, since the product of the last two terms in both traces gives a scalar integral with no loop momenta in the numerator (thanks to the cancellation against the two corresponding denominators). We have, therefore
\begin{align}
\mathbf{DB}^{(1)}_P(s,t)&=\frac{s^2}{t}\int\!\frac{d^Dk_1}{i\pi^{D/2}}\frac{d^Dk_2}{i\pi^{D/2}}\frac{\tr{k_1\,p_1p_4}\tr{k_2\,p_1p_4}}{k_1^2\left(k_1-p_{12}\right)^2k_2^2\left(k_2+p_{34}\right)^2\left(k_2+p_4\right)^2\left(k_1-k_2\right)^2\left(k_1-p_1\right)^2},\\
\label{eq:dbp2}
\mathbf{DB}^{(2)}_P(s,t)&=\frac{s\tr{p_1p_2p_4}}{t}\int\!\frac{d^Dk_1}{i\pi^{D/2}}\frac{d^Dk_2}{i\pi^{D/2}}\frac{\tr{k_2\,p_1p_4}}{\left(k_1-p_{12}\right)^2k_2^2\left(k_2+p_{34}\right)^2\left(k_2+p_4\right)^2\left(k_1-k_2\right)^2\left(k_1-p_1\right)^2},\\
\label{eq:dbp3}
\mathbf{DB}^{(3)}_P(s,t)&=\frac{s\tr{p_1p_2p_4}}{t}\int\!\frac{d^Dk_1}{i\pi^{D/2}}\frac{d^Dk_2}{i\pi^{D/2}}\frac{\tr{k_1\,p_1p_4}}{k_1^2\left(k_1-p_{12}\right)^2\left(k_2+p_{34}\right)^2\left(k_2+p_4\right)^2\left(k_1-k_2\right)^2\left(k_1-p_1\right)^2},
\end{align}
where in eqs.~(\ref{eq:dbp2}) and (\ref{eq:dbp3}) we have performed cancellations between loop momenta in the numerator and in the denominator. Recalling that the trace yields a Levi-Civita tensor, we have to compute
\begin{align}
\label{eq:dbI1}
I_{(1)}(k_1^\mu k_2^\nu)&\equiv\int\!\frac{d^Dk_1}{i\pi^{D/2}}\frac{d^Dk_2}{i\pi^{D/2}}\frac{k_1^\mu\,k_2^\nu}{k_1^2\left(k_1-p_{12}\right)^2k_2^2\left(k_2+p_{34}\right)^2\left(k_2+p_4\right)^2\left(k_1-k_2\right)^2\left(k_1-p_1\right)^2},\\
\label{eq:dbI2}
I_{(2)}{(k_2^\mu)}&\equiv\int\!\frac{d^Dk_1}{i\pi^{D/2}}\frac{d^Dk_2}{i\pi^{D/2}}\frac{k_2^\mu}{\left(k_1-p_{12}\right)^2k_2^2\left(k_2+p_{34}\right)^2\left(k_2+p_4\right)^2\left(k_1-k_2\right)^2\left(k_1-p_1\right)^2},\\
\label{eq:dbI3}
I_{(3)}{(k_1^\mu)}&\equiv\int\!\frac{d^Dk_1}{i\pi^{D/2}}\frac{d^Dk_2}{i\pi^{D/2}}\frac{k_1^\mu}{\left(k_1-p_{12}\right)^2k_2^2\left(k_2+p_{34}\right)^2\left(k_2+p_4\right)^2\left(k_1-k_2\right)^2\left(k_1-p_1\right)^2},
\end{align}
of which only terms proportional to $p_2$, $p_3$ and the metric tensor survive. Solving the momentum conservation relation for $p_3$, we can just keep terms proportional to $g^{\mu\nu}$ and $p_2^\mu$.
The integrals in eqs.~(\ref{eq:dbI2}) and (\ref{eq:dbI3}) are vector integrals already computed in Section~\ref{sec:algorithm}. They can be easily decomposed by \fare{} as follows

\begin{lstlisting}
I2=TRed[3,{k2},{{k1-p12,1},{k2,1},{k2+p34,1},{k2+p4,1},{k1-k2,1},
	{k1-p1,1}},{k1,k2}];
fI2 = I2 /. {Pair[LorentzIndex[x_], Momentum["p34"]]:>
		FourVector["p3", x] + FourVector["p4", x], 
    	     Pair[LorentzIndex[x_], Momentum["p12"]] :> 
      		FourVector["p1", x] + FourVector["p2", x]} // FullSimplify;
FI2 = fI2 /. Pair[LorentzIndex[x_], Momentum["p3"]] :> 
   		-FourVector["p1", x]-FourVector["p2", x]-FourVector["p4", x];
a2 = Coefficient[Expand[FI2], 
      Pair[LorentzIndex["m1"],Momentum["p2"]]] // FullSimplify
	@$-\mathcal{I}^5_{\{1,1,2,1,1,2\}}-\mathcal{I}^5_{\{1,1,2,1,2,1\}}
	     +\mathcal{I}^5_{\{2,1,1,1,2,1\}}-\mathcal{I}^5_{\{2,1,2,1,1,1\}}$@
	     
A2=FIREType[a2,2,{1},G] // ToExpression
	-G[{0,1,1,2,1,1,2,0,0}]-G[{0,1,1,2,1,2,1,0,0}]
	+G[{0,2,1,1,1,2,1,0,0}]-G[{0,2,1,2,1,1,1,0,0}]
\end{lstlisting}


\begin{lstlisting}
I3=TRed[3,{k1},{{k1,1}{k1-p12,1},{k2+p34,1},{k2+p4,1},{k1-k2,1},
	{k1-p1,1}},{k1,k2}];
fI3 = I3 /. {Pair[LorentzIndex[x_], Momentum["p34"]]:>
	FourVector["p3", x] + FourVector["p4", x], 
    	Pair[LorentzIndex[x_], Momentum["p12"]] :> 
      	FourVector["p1", x] + FourVector["p2", x]} // FullSimplify;
FI3 = fI3 /. Pair[LorentzIndex[x_], Momentum["p3"]] :> 
   	-FourVector["p1", x]-FourVector["p2", x]-FourVector["p4", x];

a3 = Coefficient[Expand[FI3], 
      Pair[LorentzIndex["m1"],Momentum["p2"]]] // FullSimplify
	@$\mathcal{I}^5_{\{1,1,2,1,2,1\}}-\mathcal{I}^5_{\{1,2,1,1,2,1\}}
	     -\mathcal{I}^5_{\{1,2,1,2,1,1\}}-\mathcal{I}^5_{\{1,2,2,1,1,1\}}$@
	     
A3=FIREType[a3,2,{3},G] // ToExpression
	G[{1,1,0,2,1,2,1,0,0}]-G[{1,2,0,1,1,2,1,0,0}]
	-G[{1,2,0,1,2,1,1,0,0}]-G[{1,2,0,2,1,1,1,0,0}]
\end{lstlisting}

The computation of $I_{(1)}$, that is a rank-2 tensor integral, is less trivial. Thanks to what discussed above, we are only interested in the terms proportional to the metric tensor and $p_2$. We have
\begin{lstlisting}
I1=TRed[3,{k1,k2},{{k1,1},{k1-p12,1},{k2,1},{k2+p34,1},{k2+p4,1},
	{k1-k2,1},{k1-p1,1}},{k1,k2}];
fI1 = I1 /. {Pair[LorentzIndex[x_], Momentum["p34"]]:>
		FourVector["p3", x] + FourVector["p4", x], 
    	     Pair[LorentzIndex[x_], Momentum["p12"]] :> 
      		FourVector["p1", x] + FourVector["p2", x]} // FullSimplify;
FI1 = fI1 /. Pair[LorentzIndex[x_], Momentum["p3"]] :> 
   		-FourVector["p1", x]-FourVector["p2", x]-FourVector["p4", x];
		
a1=Coefficient[Expand[FI1], MetricTensor["m1", "m2"]] // FullSimplify
	@$\mathcal{I}^5_{\{1,1,1,1,1,2,1\}}$@
A1 = FIREType[a1, 0, 2,G] // ToExpression
	-((G[{1,1,1,1,1,2,1,0,0}])/2)

(* dimension = 7 here *)	
b1 = Coefficient[Expand[FI1], Pair[LorentzIndex["m1"],Momentum["p2"]]Pair[LorentzIndex["m2"],Momentum["p2"]]] // FullSimplify;
B1=FIREType[b1,2,{0},G];
\end{lstlisting}
The expression for $B_1$ is rather lengthy and can be found in the example notebook. Nonetheless, it must be noticed that $B_1$ is the only coefficient whose integrals are in dimension 7, while for all the other quantities we have $D=5$.
It is now possibile to reduce the scalar integrals to master ones, using \fire{}. The details of the decomposition are reported in the notebook. We show here the analytical expressions of $B_1$, $A_2$, $A_3$ and $A_4$
\begin{align}
\label{eq:b1}
B_1&=\frac{2\pi}{s^2t}\left\{\frac{1}{\varepsilon}-\left[s-2t\left(1+\log4\right)+\frac{2t\log t}{\pi}\right]+\mathcal{O}(\varepsilon)\right\},\\
\label{eq:a2a3}
A_2=A_3&=\frac{2\pi}{s^2t}\left\{\frac{1}{2\varepsilon}-\left[s-t\left(1+\log4\right)+\frac{t\log t}{\pi}\right]+\mathcal{O}(\varepsilon)\right\},\\
\label{eq:a4} 
A_4&=-\frac{2\pi}{t^2}+\mathcal{O}(\varepsilon).
\end{align}
Here $A_4$ is the scalar integral obtained from the product of the two last terms in the traces of eq.~(\ref{eq:np}). As already mentioned, it does not require any tensor reduction.

It is important to highlight here a subtle feature of the dimensional reduction (DRED) employed so far. The space-time dimension of the theory imply that the external momenta are 3-dimensional vectors. However, the loop-momenta are chosen to be $D$-dimensional vectors, where $D=3-2\varepsilon$, in order to regularise the loop integrals with $\varepsilon$. This dimensional mismatch shows up in the computation of the term proportional to $A_1$. Here we have
\begin{equation}
\mathbf{DB}^{(1)}_P(s,t)=4\,\epsilon_{\mu\alpha\beta}\epsilon_{\nu\rho\sigma}\,p_1^\alpha p_1^\rho\,p_4^\beta p_4^\sigma\,I_{(1)}(k_1^\mu k_2^\nu),
\end{equation}
where $\epsilon$ denotes the Levi-Civita tensors.
As already mentioned, only the terms proportional to $g^{\mu\nu}$ and $p_2^\mu p_2^\nu$ in $I_{(1)}(k_1^\mu k_2^\nu)$ survive, so that
\begin{equation}
\mathbf{DB}^{(1)}_P(s,t)=4\epsilon_{\mu\alpha\beta}\epsilon_{\nu\rho\sigma}p_1^\alpha p_1^\rho\,p_4^\beta p_4^\sigma\,\left(A_1\,g^{\mu\nu}+B_1\,p_2^\mu p_2^\nu\right).
\end{equation}
Here, the metric tensor stemming out of DRED is implicitly $D$-dimensional, while the Levi-Civita tensors and the outer momenta are three-dimensional. The contraction of the two $\epsilon$ tensors with the $D$-dimensional metric $g^{\mu\nu}$ yields a factor $(1-2\varepsilon)$ \cite{dred}, so that the complete integral is then given by
\begin{equation}
\label{eq:dbp4}
\mathbf{DB}_P(s,t)=(1-2\varepsilon)\,s^2 t\,A_1+s^2 (s+t)\left(-s\,B_1+A_2+A_3-A_4\right).
\end{equation}
Normalisation factors are absorbed into the master integrals in $A_{1,2,3,4}$ and $B_1$. It is interesting to notice that the second term vanishes up to terms of order $\varepsilon$, i.e.
\begin{equation}
\label{eq:canc}
-s\,B_1+A_2+A_3-A_4=\mathcal{O}(\varepsilon),
\end{equation}
so that, eventually we get
\begin{align}
\label{eq:result}
\mathbf{DB}&_P(s,t)=(1-2\varepsilon)\,s^2 t\,A_1+\mathcal{O}(\varepsilon)\nonumber\\
=&-\frac{\pi}{\varepsilon^2}+\left[2(1-\log2)+\log(st)\right]\frac{\pi}{\varepsilon}+\nonumber\\
   &2\pi\left[2+\frac{2\pi^2}{3}+2\log^22+\log s\left(2\log2-1-\log t\right)\right]+\mathcal{O}(\varepsilon),
\end{align}
which matches the result found in the literature \cite{lore}.

\nocite{*}

\bibliography{fare_refs}

\end{document}